\documentclass[
reprint,
superscriptaddress,
show keys,
aps,
pra,
longbibliography,
nofootinbib
]{revtex4-2}

\usepackage{graphicx}
\usepackage{subfigure}
\usepackage{relsize}
\usepackage{xcolor}
\usepackage{mathtools}
\usepackage{amsfonts}
\usepackage{comment}
\usepackage{bbm}
\usepackage{amsmath}
\usepackage{bm}
\usepackage{cancel}

\usepackage{graphicx}
\usepackage{color}
\usepackage{colortbl}
\usepackage{array}
\usepackage{hyperref}
\usepackage[nameinlink]{cleveref}
\usepackage{mathrsfs}
\usepackage{xifthen}
\usepackage{xcolor}
\usepackage{enumitem}
\hypersetup{
	colorlinks,
	linkcolor={red!75!black},
	citecolor={blue!75!black},
	urlcolor={blue!75!black}
}
\usepackage{units}

\usepackage[utf8]{inputenc}

\newcommand{\gettitle}{Particle production and hadronization temperature in the massive Schwinger model}
\newcommand{\getHeidelbergAffiliation}{\affiliation{Institut f\" ur Theoretische Physik, Universit\" at Heidelberg, Philosophenweg 16, 69120 Heidelberg, Germany}}
\newcommand{\getJenaAffiliation}{\affiliation{Theoretisch-Physikalisches Institut, Max-Wien Platz 1, 07743 Jena, Germany}}
\hypersetup{
	colorlinks,
	linkcolor={red!75!black},
	citecolor={blue!75!black},
	urlcolor={blue!75!black},
	pdftitle={\gettitle},
	pdfauthor={Batini, },
	pdfkeywords={XXXX}
	bookmarksopen=true,
	bookmarksopenlevel=2,
	bookmarksnumbered=true
}

\allowdisplaybreaks
\begin{document}
\title{\gettitle}
\author{Laura Batini}
\email{batini@thphys.uni-heidelberg.de}
\getHeidelbergAffiliation
\author{Lara Kuhn}
\getHeidelbergAffiliation
\author{Jürgen Berges}
\getHeidelbergAffiliation
\author{Stefan Floerchinger}
\email{stefan.floerchinger@uni-jena.de}
\getJenaAffiliation
\begin{abstract}
We study the pair production, string breaking, and hadronization of a receding electron-positron pair using the bosonized version of the massive Schwinger model in quantum electrodynamics in 1+1 space-time dimensions.
Specifically, we study the dynamics of the electric field in Bjorken coordinates by splitting it into a coherent field and its Gaussian fluctuations.
We find that the electric field shows damped oscillations, reflecting pair production.
Interestingly, the computation of the asymptotic total particle density per rapidity interval for large masses can be fitted using a Boltzmann factor, where the temperature can be related to the hadronization temperature in QCD. Lastly, we discuss the possibility of an analog quantum simulation of the massive Schwinger model using ultracold atoms, explicitly matching the potential of the Schwinger model to the effective potential for the relative phase of two linearly coupled Bose-Einstein condensates.

\end{abstract}
\maketitle

\section{Introduction} \label{Introduction}
    
A long-standing puzzle in electron-positron ($e^{-}e^{+}$) collisions is that the measured hadron spectra appear approximately thermal~\cite{Castorina:2007eb}. This means that the relative abundance of two types of hadrons $a$ and $b$, of masses $m_a$ and $m_b$, respectively, is essentially determined by the ratio of their Boltzmann factors $R(a/b)$$\sim$$\exp{\{- (m_a-m_b)/T_H}\}$, with a hadronization temperature of around $T_H =160$$-$$170$ MeV. 
Surprisingly, this temperature seems to be approximately consistent across different experimental contexts, including proton-proton ($pp$), proton-antiproton ($p\bar{p}$), and heavy ion collisions~\cite{Becattini:1996gy, Cleymans:1992zc, Becattini:2008tx, Braun-Munzinger:1994ewq}.
In high-energy heavy ion collisions, a large number of partons are involved, which may lead to thermalization through rescattering. However, in $e^-e^+$ collisions, the electron and positron annihilate to produce a virtual photon that creates the quark-antiquark pair. It is unlikely that thermalization will occur through collisions in the final state. This scenario is similar to what happens in $pp$ and $p\bar{p}$ collisions.

This suggests that there is another mechanism at play, and we have previously discussed one based on entanglement between spatial regions in a QCD string \cite{Berges:2017zws, Berges:2017hne, Berges:2018cny}. Here, we want to continue this line of thought and investigate a more complete model for the quantum dynamics of hadronization.

Computing the hadronization process in the collision from first principles is challenging for several reasons. First, the process is nonperturbative~\cite{Frishman:2010tc}, meaning that it involves QCD dynamics at large distances where the coupling is strong. Moreover,  describing the process accurately requires taking into account the dynamics in time of the quark-antiquark expanding string as it breaks into hadrons during particle production.
    
Several theoretical models have been proposed~\cite{Byrnes:2002gj}, but apparently, none can account for all the observed features. The Lund model~\cite{Andersson:1983ia, Andersson:1997xwk} is based on expanding QCD strings from which tunneling processes produce hadrons and resonances via the Schwinger mechanism~\cite{Gelis:2015kya}. The standard implementation of the Lund model is the PYTHIA event generator~\cite{Sjostrand:2006za, Sjostrand:2007gs, Sjostrand:2014zea, Fischer:2016zzs} for which it seems difficult to explain the thermal-like features seen in experimental data without additional modifications~\cite{Fischer:2016zzs}.
 
As explained above, the full problem of particle production in QCD is challenging. Therefore, one can resort to effective models that are simpler to study and can still explain several aspects of the original model.
The Schwinger model is quantum electrodynamics in 1+1 space-time dimensions (QED$_{1+1}$)~\cite{Schwinger:1962tp, Coleman:1976uz, LOWENSTEIN1971172, Jayewardena:1988td}.  Despite being a $U(1)$ Abelian gauge theory, it shares several essential features with QCD, which is a $SU(3)$ gauge theory, making it an effective toy model for simulating some QCD characteristics~\cite{Jentsch:2021trr,Grieninger:2023ehb}. QED$_{1+1}$ includes the Higgs mechanism, charge screening, ``quark'' confinement~\cite{Coleman:1975pw, Gross:1995bp}, string breaking~\cite{Andersson:1983ia, Antonov:2003ir, Hebenstreit:2013baa, Hebenstreit:2013qxa, Hebenstreit:2014rha,  Spitz:2018eps, Florio:2023dke, Florio:2024aix}, and spontaneous chiral symmetry breaking, as well as topological vacua.
Furthermore, transverse and longitudinal degrees of freedom naturally separate at high energies, thus justifying the use of a dimensionally reduced theory.
    
Another advantage of this theory is that the fermionic theory of QED$_{1+1}$ can be rewritten in terms of a bosonic theory with a real scalar field in $d=1+1$~\cite{Hamer:1982mx, Berges:2017hne, Berges:2018cny}. The simplicity of the resulting scalar theory makes it suitable to study real-time dynamics. 
This model, in its bosonic formulation, has proved to be successful for the study of the anomalous photon production puzzle, predicting the enhancement of soft photon production in agreement with the experimental results~\cite{Kharzeev:2013wra, Loshaj:2014aia}. Moreover,  qualitative features of jet fragmentation have been reproduced~\cite{PhysRevD.10.732}. 

This work aims to study whether the hadronization temperature can be recovered in QED$_{1+1}$ in its bosonic version and whether it shows qualitative and quantitative agreement with the experimental results.
To do so, we simulate the dynamics of an electron-positron pair flying apart to study pair production during the string fragmentation process. We analyze the spectrum of the produced particles, which allows for direct computation of the Boltzmann factor that can be quantitatively compared with experimental results.
Additionally, we observe the phenomenon of coherent damped field oscillations induced by the propagating high-energy quark due to the continuous creation of quark-antiquark pairs pulled from the vacuum to screen the electric charges.
This finding aligns with previous studies that have demonstrated similar results using lattice simulations~\cite{Spitz:2018eps, Hebenstreit:2013baa, Hebenstreit:2014rha}.

Recently, there has been much interest in low-dimensional dynamics with the ultimate goal of simulating gauge theories using quantum simulators~\cite{Banerjee:2012pg,Grieninger:2024axp,Dalmonte:2016alw,Goldman:2014xja}.
Various works have proposed the analog simulation of the massive Schwinger model using ultracold quantum gases~\cite{Wiese:2013uua,  Kasper:2016mzj, Zohar:2011cw, Zohar:2012ay, Kasper:2015cca}, and the quantum simulation of its nonperturbative aspects using tensor networks~\cite{Kuhn:2014rha}, and spin chains~\cite{Verdel:2019chj}.
We explore a possible experimental realization of this model using ultracold atoms, in particular, two tunnel-coupled Bose-Einstein condensates (BEC) with a modulated linear interaction.

The paper is organized as follows. In Sec.~\ref{sec:QED1+1}, we introduce the model of QED in 1+1 dimensions, its properties and similarities with QCD, and how to formulate its bosonized version. 
In Sec.~\ref{sec:ExpandingStringsDynamics}, we study the dynamics of the expanding string. We discuss in Sec. \ref{sec:Falsevacuum} the case of metastable initial conditions for the field and the resulting time evolution.
Section \ref{sec:ParticleProduction} introduces Gaussian fluctuations to an underlying oscillating background electric field and employs the Bogoliubov coefficients~\cite{Davies} method to compute the produced particles' spectra. We establish the emergence of a temperature as the Boltzmann factor. Section~\ref{sec:AnalogQuantumSimulation} introduces an experimental proposal to study the massive Schwinger model using an ultracold atom system. In Sec.~\ref{sec:Conclusions}, we conclude and give an outlook. 

Throughout this work, we use units where $\hbar=c=k_\text{B}=1$. Our metric in four dimensions is mainly plus, so in two dimensions it is $\eta_{\mu\nu}=\text{diag}(-1,+1)$, and the Levi-Civita tensor is such that $\epsilon^{01}=-\epsilon^{10}=-\epsilon_{01}=\epsilon_{10}=1$.

\section{QED in 1+1 space-time dimensions} \label{sec:QED1+1}
As explained and motivated in the introduction, we assume that QED$_{1+1}$ can model relevant aspects of the dynamics for the production of quark-antiquark pairs in the string fragmentation. 
In this section, we first introduce the microscopic QED$_{1+1}$ model, then discuss the bosonization procedure and its implications.
    
\subsection{Gauge theory action}
For QED in $d=1+1$ dimensions with a single massive fermion $\psi$, the microscopic action is
\begin{equation}
	S= \int \mathrm{d}t \mathrm{d}x\left\{ -\frac{1}{4} F_{\mu \nu} F^{\mu \nu}-\bar{\psi}\gamma^\mu (\partial_\mu - i q A_\mu) \psi - m \bar \psi \psi \right\},
    \label{eq:QED1+1Lagrangian}
\end{equation}
with $U(1)$ gauge field $A_\mu(t, x)$ and one fermion flavor $\psi(t, x)$. The free parameters are given by the fermion mass $m$ and electric charge $q$, which both have dimensions of mass. This makes the model super-renormalizable so that it can be defined without an explicit dependence on an ultraviolet regularization.

The electromagnetic field strength tensor has only one independent component
\begin{equation}
    F_{\mu \nu}=\partial_\mu A_\nu-\partial_\mu A_\nu = \epsilon_{\mu \nu} F_{10},
\end{equation} 
identified as the electric field $E=F_{10}$. Note that there is no magnetic field.

  Furthermore, the Schwinger model features topological $\theta$-vacua similar to QCD in 1+3 dimensions, see below. 

\subsection{Bosonized action} \label{app:bosonization}
The Schwinger model has an alternative bosonized description in terms of a real scalar field $\phi(t, x)$, which is linearly related to the electric field,
\begin{equation}
    E(t,x)=\frac{q\phi(t,x)}{\sqrt{\pi}}.
\label{eq:electricfield}
\end{equation}

We discuss the bosonized formulation and its renormalization in Appendix~\ref{app:renormalization_Schwinger_model}. It is shown there that a renormalized effective action can be written in the form 
\begin{equation}
    \Gamma[\phi] = \int \mathrm{d}^2 x \sqrt{g} \left\{- \frac{1}{2} g^{\mu\nu} \partial_\mu \phi \partial_\nu\phi - V(\phi) \right\}.
\label{eq:renormalizedaction}
\end{equation}
For later convenience, we have allowed general coordinates and introduced the corresponding metric $g_{\mu\nu}$ with $g=-\det(g_{\mu\nu})$. The potential is 
\begin{equation}
    V(\phi) = \frac{1}{2} M^2 \phi^2 -  u \cos\left( 2\sqrt{\pi} \phi + \theta \right),
\label{eq:renormalizedPotential}
\end{equation}
where
\begin{equation}
    M = \frac{q}{\sqrt{\pi}}\, , \quad\quad u = \frac{\exp(\gamma) q m}{2\pi^{3/2}}\, .
\end{equation}
The potential in Eq.\ \eqref{eq:renormalizedPotential} was also found in Ref.\ \cite{Coleman:1975pw}. 
We note that it is a renormalized potential in the sense that some quantum fluctuations have been taken into account through the renormalization of $u$. However, it cannot be seen as the effective potential corresponding to full quantum effective action, which would have to be convex as a Legendre transform. 

In the following, we also work with the dimensionless ratio
\begin{equation}
    \kappa = \frac{2\sqrt{\pi} u}{M^2} = \frac{\exp(\gamma) m}{q}
    \label{eq:dimensionlesscoupling}
\end{equation}
between the cosine and quadratic term in the potential. 

\subsection{Potential}
\begin{figure*}[t]
    \centering 
    \includegraphics[width=.49\textwidth]{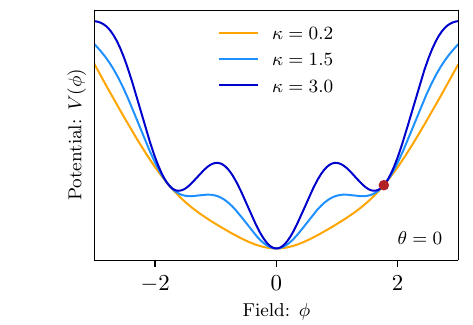}
    \includegraphics[width=.49\textwidth]{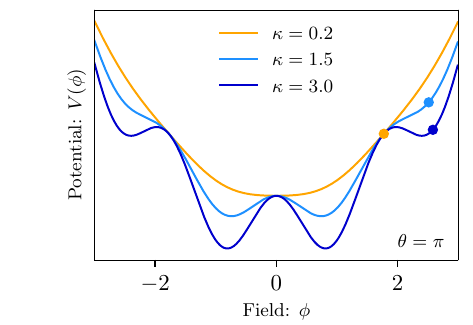}
    \caption{Potential of the massive Schwinger model $V(\phi)$ [Eq.~\eqref{eq:renormalizedPotential}] for $\theta = 0$ (left panel) and $\theta = \pi$ (right panel). The different colors represent different dimensionless coupling strengths $\kappa$ [Eq.~\eqref{eq:dimensionlesscoupling}]. The dots indicate the initial condition for the field according to Eq.~\eqref{eq:initialconditions}. }
        \label{fig:classical potential}
\end{figure*}

The minima of the potential $V(\phi)$ in Eq.~\eqref{eq:renormalizedPotential} can be identified with vacuum states of the quantum theory. In particular, the ground state is determined by the field $\Phi_{\mathrm{vac}}$ at the global minimum of the potential. Its position depends on the vacuum angle $\theta$.

In addition to the global minimum, there are also local minima corresponding to metastable states~\cite{Ai:2020vhx}, where the field can remain trapped for a while but eventually relaxes by quantum tunneling~\cite{Coleman:1977py, Coleman:1977th}.

The potential $V(\phi)$ is displayed for two possible values of the vacuum angle $\theta = 0$, and $\theta= \pi$ in Fig.~\labelcref{fig:classical potential}.
For $\theta = \pi$ there are two degenerate minima, with a $Z_2$ symmetry that can break spontaneously. For $\theta=0$, the potential has a unique minimum for any choice of $q$ and $m$.

In the strong interaction limit $q^2 \gg m^2$, or $u \ll M^2$, one recovers the massless Schwinger model, equivalent to a free scalar field theory with a boson of mass $M$ with one single minimum of the potential at $\phi=0$.

\subsection{Weak coupling limit}
In the weak coupling limit $m\gg q$, the quadratic term $\sim$~$M^2$ in the potential in Eq.~\eqref{eq:renormalizedPotential} [or, equivalently for the microscopic potential in Eq.~\eqref{eq:bosonicpotential}] is subleading compared to the cosine term and one has effectively a kind of sine-Gordon theory~\cite{Coleman:1974bu}. It is instructive to discuss this limit where intuition from perturbative QED can be applied.

The potential has for $M^2 \ll u$ a set of almost degenerate minima at $2\sqrt{\pi}\phi+\theta = 2\pi n$, with $n\in\mathbb{Z}$. The corresponding electric fields differ according to Eq.~\eqref{eq:electricfield} by multiples of the charge $q$. One can see these field configurations as being due to an integer number of elementary charges $q$, and in fact, the original charged fermions correspond in the bosonized theory to solitons that connect neighboring minima of the potential.

For the particular case of $\theta = 0$, the potential has its global minimum at $\phi=0$, and the field configuration with $\phi_1 \approx \pm \sqrt{\pi}$ would be the next-to-lowest state. Its energy per unit length with respect to the ground state is given by
\begin{equation}
    \sigma = \frac{1}{2} M^2 \phi_1^2 = \frac{1}{2} q^2.
\label{eq:fieldenergy_QED1+1}
\end{equation}
This can also be understood as the energy density of the electric field $E^2/2$. Note that in $d=1+1$, the energy stored in the field grows linear with length, i.e., with the separation between two opposite charges, and $\sigma$ has the physical significance of a string tension.

When the term $\sim M^2$ gains relevance compared to the term $\sim u$, the string tension increases, and the corresponding field configurations become more unstable. The string can then break by quantum tunneling or false vacuum decay in the bosonic formulation. In the original fermionic formulation, this corresponds to the creation of quark-antiquark pairs. We are interested in studying the dynamics of this process in further detail.

\section{Dynamics of expanding strings} \label{sec:ExpandingStringsDynamics}

Let us now turn to the problem of an expanding QCD string in the reduced model, where the dynamics happens in $d=1+1$ dimensions and massive QED is employed instead of QCD. We start from a quark-antiquark pair with very high energy, where the quark and antiquark move in opposite directions with the speed of light. The electric field that forms between them will be modeled as a coherent scalar field in the bosonized description and we will study its evolution, as well as the excitations around it. 

\subsection{Coordinate system}

We start by introducing a particularly convenient coordinate system to describe the string resulting from a highly energetic quark-antiquark pair. Without loss of generality, the string is taken to be confined to the $x$-direction, and the trajectories of the moving quark-antiquark pair on the light cone are $x = \pm t$, $y=z=0$. 

This geometry can be conveniently described using Bjorken coordinates $(\tau, \eta)$, where $\tau$ stands for proper time, and $\eta$ indicates the space-time rapidity. These coordinates are related to Minkowski coordinates $(t,x)$ via $\tau=\sqrt{t^2-x^2}$ and~$\eta=\operatorname{artanh}(x / t)$ such that $t=\tau \cosh \eta$ and $x=\tau \sinh \eta$. The invariant line element is in Bjorken coordinates given by $\mathrm{d}s^2 = - \mathrm{d}\tau^2 + \tau^2 \mathrm{d} \eta^2$. Standard Bjorken coordinates are defined for $t \geq|x|$, i.e., on and within the future light cone of the origin, but a similar construction would work for the past light cone. The light cone itself corresponds to $\eta\to \pm \infty$ or $\tau = 0$. 

Expressed in Bjorken coordinates, the equation of motion for the scalar field as obtained from the variation of Eq.~\eqref{eq:renormalizedaction} reads [using the ratio in Eq.~\eqref{eq:dimensionlesscoupling}]
\begin{equation}
\begin{aligned}
    & \partial_\tau^2 \phi(\tau, \eta)+\frac{1}{\tau^2} \partial_\eta^2 \phi(\tau, \eta)+\frac{1}{\tau} \partial_\tau \phi(\tau, \eta)  \\ & + M^2 \left[ \phi(\tau, \eta) + \kappa \sin (2 \sqrt{\pi} \phi(\tau, \eta)+\theta) \right] = 0.
\end{aligned}
\label{eq:EOMfield}
\end{equation}

Bjorken coordinates are particularly useful because of the symmetry of the string expansion with respect to longitudinal boosts, $\eta \to \eta + \Delta \eta$. The field expectation value can be assumed to be invariant under this boost transformation realized as a translation. For quantum excitations the symmetry is realized in a statistical sense.

\subsection{Background field evolution}\label{sec:BackgroundFieldEvolution}

We solve the dynamics of string breaking in several steps and start with the simplest case of a coherent or classical field. 

As a consequence of Bjorken boost symmetry~\cite{Floerchinger:2011pxy}, the evolving field expectation or background field does not depend on the rapidity variable, $\phi(\tau, \eta)=\Phi(\tau)$. The equation of motion \eqref{eq:EOMfield} reduces to the evolution equation
\begin{equation}
    \partial_\tau^2 \Phi+\frac{\partial_\tau \Phi}{\tau} +M^2 \left[ \Phi + \kappa \sin (2 \sqrt{\pi} \Phi+\theta) \right] = 0 .
    \label{eq:coherentbackgroundfield}
\end{equation}
\begin{figure*}
\includegraphics[width=.49\textwidth]{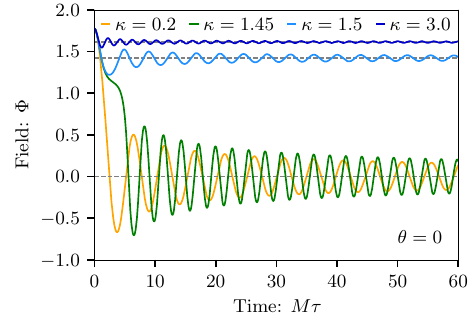}
\includegraphics[width=.49\textwidth]{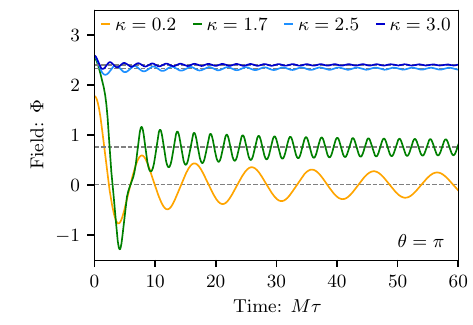}
\caption{Background field for vacuum angles $\theta = 0$ (left panel) and $\theta = \pi$ (right panel) with different couplings $\kappa$. The dashed lines indicate the position of the minimum around which $\Phi$ oscillates for $M \tau \gg 1$.}
\label{fig:backgroungfield}
\end{figure*}
The initial condition $\Phi(\tau_0)$ with $\tau_0= 0^+$ is fixed by the requirement that the classical electric field of two relativistic charges $\pm q$ flying in opposite directions is directly on the light cone given by $E=q$. Quantum effects are assumed to be negligible in this early time limit, see also Ref.~\cite{Iso:1988zi}. Thus, the initial value of the coherent scalar field can be obtained from the relation~\eqref{eq:electricfield} as
\begin{equation}
    \Phi(\tau_0)=\Phi_{\mathrm{vac}}+\sqrt{\pi}.
    \label{eq:initialconditions}
\end{equation}
In Fig.~\ref{fig:classical potential}, we denote the initial field value for a number of parameters by dots.

It is instructive to discuss first a linearized form of Eq.~\eqref{eq:coherentbackgroundfield}, where the linearization is done around the initial state \eqref{eq:initialconditions},
\begin{equation}
    \partial_\tau^2 \Phi+\frac{\partial_\tau \Phi}{\tau} +M^{\prime 2} \Phi = I,
    \label{eq:coherentbackgroundfieldlinear}
\end{equation}
where 
\begin{equation}
M^{\prime 2} = M^{2}\left[ 1 + 2 \sqrt{\pi} \kappa \cos (2 \sqrt{\pi} \Phi_\text{vac}+\theta) \right]
\label{eq:modM2}
\end{equation}
and the inhomogeneous term is
\begin{equation}
\begin{split}
    I = & M^{2} \kappa \left[ (2\sqrt{\pi} \Phi_\text{vac}+2\pi)\cos(2\sqrt{\pi} \Phi_\text{vac}+\theta) \right. \\
    & \left.-  M^{2} \kappa \sin(2\sqrt{\pi} \Phi_\text{vac}+\theta) \right].    
\end{split}
\end{equation}
The homogeneous equation has the form of Bessel's differential equation, while a particular solution of the inhomogeneous equation is the constant $\Phi = I/M^{\prime 2}$. Accordingly, the solution can be written as a linear combination of Bessel functions of the first and second kind,
\begin{equation}
    \Phi(\tau)=I/M^{\prime 2} + c_1 J_0(M^\prime \tau)+c_2 Y_0(M^\prime \tau) \, .
\end{equation}
The coefficients $c_1$ and $c_2$ are determined through the initial condition. Specifically, the Bessel function of the first kind becomes unity, $J_0(M^\prime \tau) \rightarrow 1$, for $\tau \rightarrow 0$, while the Bessel function of the second kind diverges in that limit. This sets $c_1=\Phi_\text{vac}+\sqrt{\pi}-I/M^{\prime 2}$ and $c_2=0$ and we arrive at the early time solution
\begin{equation}
\Phi_{\text{in}}(\tau)= I/M^{\prime 2} + \left[\Phi_\text{vac}+\sqrt{\pi}-I/M^{\prime 2} \right] J_0(M^\prime \tau).
\end{equation}
In order to fully understand the evolution of the coherent field $\Phi(\tau)$ also at late times, we numerically solve the nonlinear equation of motion, Eq.~\eqref{eq:coherentbackgroundfield}, using the fourth-order Runge-Kutta method. Figure~\ref{fig:backgroungfield} illustrates the behavior of $\Phi(\tau)$ for various choices of the ratio $\kappa$ and the vacuum angles $\theta=0$ as well as $\theta = \pi$.

Starting from the initial value, the field rolls down the potential and eventually oscillates around one of its minima. In addition, the overall expansion implies a dilution and an effective damping, similar to the Hubble damping for a scalar field in the early Universe. In the asymptotic limit $\tau\to \infty$, the field expectation value approaches a local minimum according to the classical evolution.

Note that depending on the value of $\kappa$, there are qualitative differences in the dynamics of the background field. For very small values $\kappa\ll 1$, corresponding essentially to small fermion mass $m\ll q$, the potential has only a single global minimum which is then approached by the classical evolution equation in the asymptotic long time limit. In contrast, when $\kappa$ is larger, the background field can be trapped in a local minimum, which corresponds to a metastable state.

In Sec.~\ref{sec:ParticleProduction}, we will discuss the production of quantum excitations around the coherent background field solution $\Phi(\tau)$. However, before that, we first address the metastable background solutions and the decay of such false vacuum states through field tunneling.

\section{False vacuum decay} \label{sec:Falsevacuum}

\begin{figure*}
\centering
\includegraphics[width=.49\textwidth]{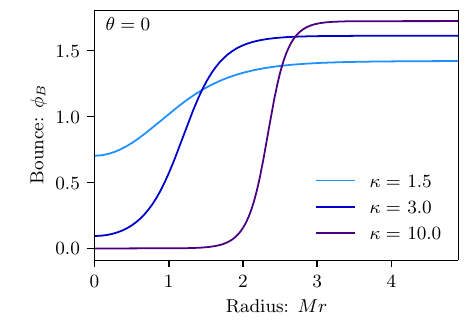} 
\includegraphics[width=.49\textwidth]{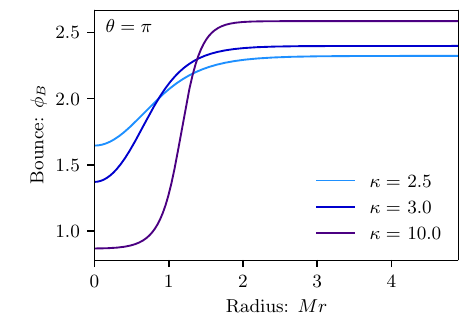} 
\includegraphics[width=.49\textwidth]{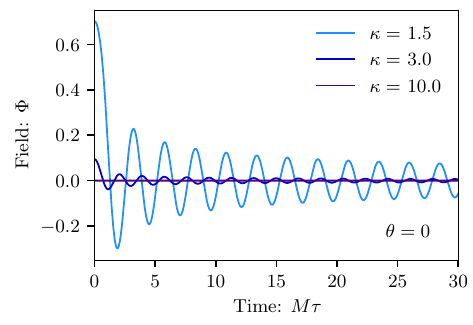}
\includegraphics[width=.49\textwidth]{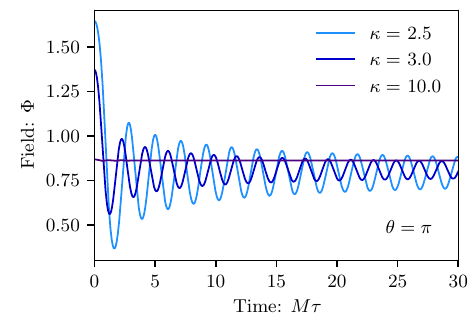}
\caption{\textbf{Upper}: Bounce solutions $\phi_B$ in dependence on the radial coordinate r and for vacuum angles $\theta = 0$ (left panel) and $\theta = \pi$ (right panel) and various couplings $\kappa$. \textbf{Lower}: Evolution of the background field $\Phi$ after tunneling. The background field $\Phi$ is displayed in dependence on Bjorken time $\tau$ for $\theta = 0$ (left panel) and $\theta = \pi$ (right panel) with various couplings $\kappa$.}
\label{fig:bounceFVD}
\end{figure*}

The potential of Eq.~\eqref{eq:renormalizedPotential}, as plotted in Fig.\ \ref{fig:classical potential}, shows for larger values of $\kappa$ local minima in addition to the global minimum (or two global minima for $\theta = \pi$). These correspond to metastable states in the full quantum theory. 

The decay of such a metastable configuration has close parallels to (QCD) string breaking. One possibility is that the field $\phi$ gets trapped during its evolution in such a local minimum and needs quantum tunneling to evolve further. But one can also see the metastable state directly as an approximation to the string configuration and its decay by a tunneling event as being the analog of the spontaneous production of a quark-antiquark pair. Moreover, we will see that a tunneling event at some spacetime position $x^\mu_0$ leads, within the future light cone of this spacetime point, to an evolving field that resembles very much the evolving background field produced by a highly energetic quark-antiquark pair.

Let us start our discussions from a field in a metastable local minimum configuration that we denote by $\phi_{FV}$. This would be a stable solution of the equations of motion in a classical sense, but in the full quantum theory, the state is only metastable as tunneling through the barrier can occur, allowing the field to reach the true vacuum eventually (that we denote by $\phi_{TV}$). In the case of $\theta=\pi$, $\phi_{TV} \approx \pm \sqrt{\pi} / 2$, while for  $\theta=0$, $\phi_{TV}=0$.

The decay of false vacua usually occurs through nucleation and expansion of spherical bubbles in space~\cite{Coleman:1977py, Coleman:1977th, Stone:1975bd}, a process common to first-order phase transitions across different fields such as condensed matter~\cite{LANGER1967108}, particle physics~\cite{Arnold:1989cb}, and cosmology~\cite{Markkanen:2018pdo}.
In this work, we only focus on the nucleation of a single bubble that expands and do not consider the possibility of multiple bubbles forming and colliding~\cite{Freivogel:2009it, Hawking:1982ga}. 
This bubble solution, also known as the critical bubble, bounce or instanton~\cite{etde_6212841}, denoted by $\phi_B$, depends on a radial Euclidean coordinate $r^2 = t_{\textrm{E}}^2+ x^2$,  where $t_{\textrm{E}}$ denotes imaginary or Euclidean time. The coordinate origin corresponds here roughly to the space-time point where the string starts to break. In Euclidean coordinates the critical bubble solution has the boundary conditions $\phi_B(r\to \infty) = \phi_{FV}$, and some field value, $\phi_B(r = 0) = \phi_0$, which is to be determined at vanishing radius, and the first derivative is supposed to vanish there, $\partial_r \phi_B(r=0) = 0$. For more details on computing this field configuration, we refer to Appendix~\ref{App:FalseVacuumDecay}.

The upper left and right panels of Fig.~\ref{fig:bounceFVD} show bounce profiles in dependence on the radial coordinate $r$ for $\theta = 0$ and $\theta=\pi$, respectively. As the coupling $\kappa$ increases, the value at small Euclidean radius $\phi_B(0)=\phi_0$ shifts towards the true vacuum $\phi_{TV}$. Moreover, the steepness of the bounce changes with $\kappa$. The bounce resembles a step function for large values of $\kappa$ (e.g., $\kappa=10.0$). In this case, we are in the thin-wall approximation, and the position of the step can be identified as the bubble radius $R$ that separates the region filled with the true vacuum phase and the false vacuum background.

It is an interesting problem to follow the time evolution in real time within the future light cone of the breaking point at the coordinate origin. This leads to a very similar problem to the one discussed in the previous section. Directly on the light cone, initial conditions are set through the bounce solution, but everything in the future of that is to be computed dynamically.

The usage of Bjorken coordinates is beneficial, where the boundary conditions for the time evolution of a coherent background field in Euclidean coordinates can be rewritten as the initial conditions
\begin{equation}
    \Phi(\tau=0) = \phi_B(0)=\phi_0, \quad \partial_\tau \Phi(\tau=0) =0.
\label{eq:initialFluctuationsAfterBounce}
\end{equation}
Displayed on the lower plots of Fig.~\ref{fig:bounceFVD}, we show the resulting background field $\Phi$ evolution for the new initial conditions (for $\theta = 0$ in the left panel and $\theta = \pi$ in the right panel). Again, the field shows damped oscillations, now around the true vacuum.
    
Outside of the future and past light cones of the origin, the shape of the bubble in Minkowski space can be determined by analytically continuing the bounce solution~\cite{Coleman:1977py, Bergner:2002we}. As the bounce is spherically symmetric in Euclidean space, its analytic continuation is $O(1, 1)$ symmetric,
\begin{equation}
\phi_B\left(r=\sqrt{t_E^2+x^2}\right)=\phi_B\left(\sqrt{x^2-t^2}\right). 
\end{equation}
This continuation is indeed only possible for $|x| > |t|$. Otherwise, the radicand would become negative, and the radius $r$ would be imaginary.

\begin{figure}
    \centering
\includegraphics[width=.5\textwidth]{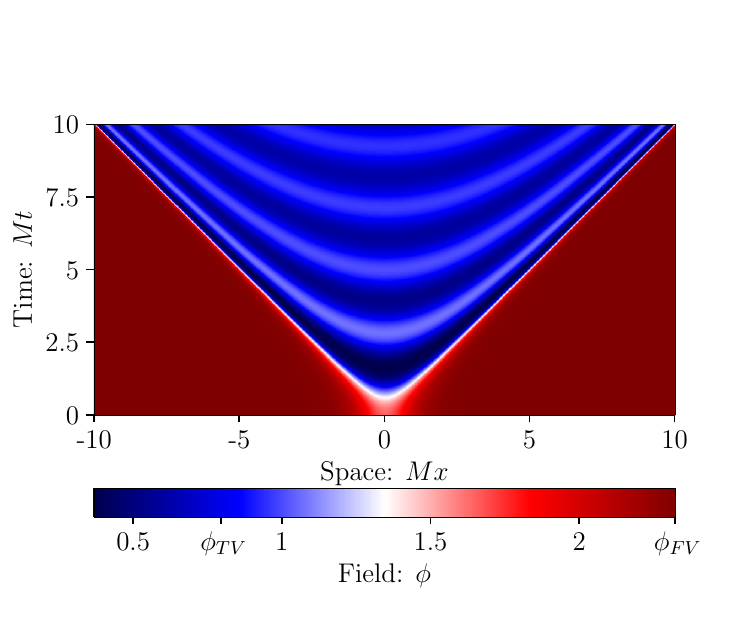}
    \caption{Field evolution in Minkowski coordinates $(x,t)$. Outside the forward light cone, i.e., for $|x|>t$, we show the analytic continuation of the bounce solution $\phi_B$. Inside the light cone, i.e., for $t>|x|$, the solution is obtained by solving the classical equations of motion. We have chosen here $\kappa = 2.5$ and $ \theta =\pi$. The color coding indicates the field values, with $\phi_{TV}$ being the true vacuum and $\phi_{FV}$ the initial false vacuum.}
    \label{fig:light cone}
\end{figure}

Combining both solutions, the analytically continued bounce solution for $|x|>t$ and the solution to classical evolution equations for $t>|x|$ is shown in Fig.~\ref{fig:light cone}. The field decays from the false vacuum outside the light cone to the true one inside it, showing damped oscillations very similar to those obtained in Sec.~\ref{sec:BackgroundFieldEvolution}.

\section{Particle production} \label{sec:ParticleProduction}
Now that we have obtained nontrivial evolving background field configurations in two closely related scenarios, it is interesting to study quantum fluctuations around them. Field excitations can be seen as particles, and in this section, we aim to compute the resulting particle production~\cite{Felder:2001kt, Abolhasani:2009nb, Dufaux:2006ee,Kofman:1997yn}. 
\subsection{Adding quantum fluctuations}
\begin{figure*}
    \centering        
\includegraphics[width=.49\textwidth]{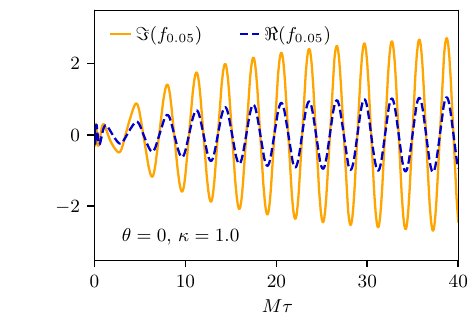}
\includegraphics[width=.49\textwidth]{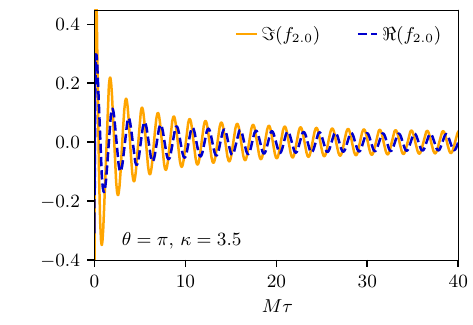}
    \caption{Real part (solid line) and imaginary part (dashed line) of the mode function evolution in time. We show three choices of the coupling ratio $\kappa$ defined in Eq.~\eqref{eq:dimensionlesscoupling} and the rapidity wave number $k$.}
        \label{fig:modefunctions}
\end{figure*}
We add a quantum fluctuation field $\varphi$ to the coherent classical background field $\Phi$ such that $ \phi(\tau,\eta)=\Phi(\tau)+\varphi(\tau, \eta)$.

For this quantum field, we employ the mode expansion
\begin{equation}
    \varphi(\tau, \eta)=\int \frac{\mathrm{d} k}{2 \pi}\Big\{a_k f_k(\tau) e^{ik\eta}+a^{\dagger}_k f_k^*(\tau) e^{-ik\eta}\Big\},
\label{eq:fieldphimodeexpansion}
\end{equation}
with mode functions $f_k(\tau)$ that satisfy the differential equation
\begin{equation}
\begin{split}
& {\bigg (} \partial_\tau^2+\frac{\partial_\tau}{\tau} +\frac{k^2}{\tau^2} \\
& +M^2 \left[ 1 +2 \sqrt{\pi} \kappa \cos (2 \sqrt{\pi} \Phi(\tau)+\theta) \right] {\bigg )} f_k(\tau)=0.
\end{split}\label{eq:modefunctions}
\end{equation}
This equation is obtained by linearising Eq.~\eqref{eq:EOMfield} around the background solution $\Phi(\tau)$.

The $a_k$ and $a^{\dagger}_k$ are annihilation and creation operators, respectively, which obey the commutation relations $[a_k, a^{\dagger}_{k^{\prime}}]=2 \pi \delta(k-k^{\prime})$ and $[a_k, a_{k^{\prime}}]=0$.
We impose the normalization condition
\begin{equation}
    -i f_k(\tau) \tau \partial_\tau f_k^*(\tau)+i f_k^*(\tau) \tau \partial_\tau f_k(\tau)=1,
\end{equation}
such that the equal-time canonical commutation relation reads with the conjugate momentum field $\Pi(\tau, \eta)=\delta \Gamma / \delta(\partial_\tau \phi(\tau, \eta)) = \tau \partial_\tau \phi(\tau, \eta)$,
\begin{equation}
    [\phi(\tau, \eta), \Pi(\tau, \eta^\prime)] = i \delta(\eta-\eta^\prime). 
\end{equation}

    \subsection{Initial choice of mode functions}
We need to find proper initial conditions to solve the equations of motion for the mode functions. 
It is useful to discuss here first the case where the mean field is in the ground state, so constant and positioned at the global minimum of the potential, $\Phi(\tau) = \Phi_\text{vac}$. In that case, an analytic solution of Eq.~\eqref{eq:modefunctions} becomes possible in terms of Hankel functions of the second kind,
\begin{equation}
    f_k^{\text{init}}(\tau_0)=\frac{\sqrt{\pi}}{2} e^{\frac{\pi}{2}k} H_{i k}^{(2)}(M^\prime \tau_0),
\label{eq:Hankelfunctions}
\end{equation} 
where $M^\prime$ is given in 
Eq.\ \eqref{eq:modM2}.
At first sight, this choice does not seem to be unique because a solution of the mode equation~\eqref{eq:modefunctions} could also be found, for example, in terms of Bessel functions of the first kind. However, one can argue that the Hankel function solution in Eq.~\eqref{eq:Hankelfunctions} is the right choice. 
To see this, consider the integral representation
\begin{equation}
    H^{(2)}_{i k}(M^\prime \tau) = \frac{i}{\pi}e^{-\frac{\pi}{2}k}\int_{- \infty}^{+ \infty} \mathrm{d}z e^{-iM^\prime \tau \cosh(z) - i k z}.
\end{equation}
After a shift of the integration variable $z \to z+\eta$ and returning to Minkowski coordinates $t$ and $x$ using $\tau \cosh(z+\eta) = \tau \cosh(z) \cosh(\eta) + \tau \sinh(z) \sinh(\eta) = \cosh(z) t + \sinh(z) x$ we obtain
\begin{equation}
    f^\text{init}_k(\tau) e^{ik\eta} = \frac{i}{2 \sqrt{\pi}} \int_{- \infty}^{+ \infty} \mathrm{d}z   e^{-iM^\prime [\cosh(z) t + \sinh (z) x] -ik z} . 
\end{equation}
One can see here that the modes described by $f^\text{init}_k(\tau)$ are superpositions of plane waves with positive frequencies $\omega(z) = M^\prime \cosh(z)$ in Minkowski space. 

To the initial mode functions in Eq.\ \eqref{eq:Hankelfunctions} correspond operators $a^\text{init}_k$ and a unique quantum state $|\Omega, \Phi_\text{vac} \rangle$ annihilated by them,  $a^\text{init}_k | \Omega, \Phi_\text{vac} \rangle = 0$. This state represents the vacuum state within the here employed parametrization of a coherent background field with small excitations around it.

Another interesting state to be considered is a coherent state that basically agrees with the vacuum state, but where the expectation value is shifted, $\Phi_\text{vac}\to \Phi(\tau_0)$. We denote this state by $|\Omega, \Phi(\tau_0) \rangle$ with $\Phi(\tau_0)$ the field expectation value $\Phi(\tau)$ at the initial time $\tau=\tau_0$. For the scenario discussed in Sec.~\ref{sec:ExpandingStringsDynamics} this initial value is given in Eq.\ \eqref{eq:initialconditions}, while for the scenario discussed in Sec.~\ref{sec:Falsevacuum} it is specified in Eq.~\eqref{eq:initialFluctuationsAfterBounce}. For the mode functions in this state, we use at very early times $\tau\to 0^+$ the form given in Eq.~\eqref{eq:Hankelfunctions} without any modification. Of course, deviations will occur at later times $\tau > 0$, see below.

The ``shifted vacuum'' coherent state specified through this prescription is only one out of many possible quantum states one could consider at $\tau = 0^+$. In particular for the bounce scenario discussed in Sec.~\ref{sec:Falsevacuum} one may expect that one could do better and actually determine the quantum state of fluctuations after the bounce possibly through analytic continuation from a calculation in Euclidean space, but we leave this for future work.

\begin{figure*}
    \centering
 \includegraphics[width=.49\textwidth]    {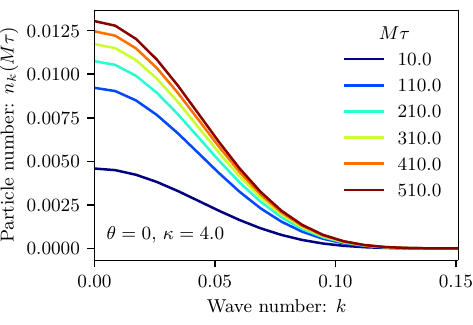}   
    \includegraphics[width=.49\textwidth]{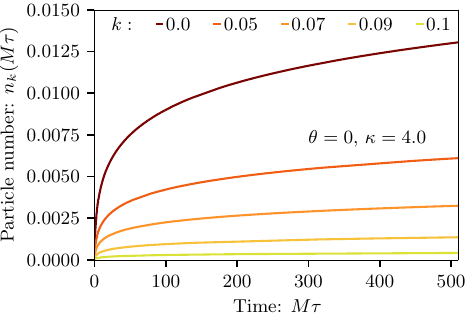}
    \caption{Particle number $n_k$ dependence on the rapidity wave number $k$ (left panel) and the corresponding dependence on time $M\tau$ (right panel) for a selection of rapidity wave numbers $k$ as shown in the figure.}
    \label{fig:modefunctions2}
\end{figure*}

Starting with the initial conditions \eqref{eq:Hankelfunctions} for very early times, we use the differential equations \eqref{eq:modefunctions} to find solutions for later times. 
Figure~\ref{fig:modefunctions} shows the real and imaginary part of the mode functions obtained in this way for different choices of the coupling constant $\kappa$ and the rapidity wave number $k$. In the right panel of Fig.~\ref{fig:modefunctions2}, we show the growth in time of the occupation numbers [see Eq.\ \eqref{eq:occnumbfinal} below for their definition] for several wave number $k$ for $\theta = 0, \kappa = 4.0$. 

In general, the linearization procedure and the Bogoliubov theory are valid as long as the effect of the fluctuations effect on the background field is small and if the fluctuations remain small during the evolution. Due to the effect of the Bjorken expansion, the occupation numbers exhibit exponential growth only for a time window, after which their value saturates. This phenomenon is discussed in detail in Ref.~\cite{Berges:2012iw}.
Even though the backreaction of the fluctuations is expected to introduce some extra damping in the background field dynamics, it does not significantly alter the qualitative dynamics governing the mode function equation. This analysis is left for future work.

\subsection{Asymptotic particle number}
As a consequence of the overall expansion and dilution of energy, the background field $\Phi(\tau)$ eventually approaches a constant asymptotic value for very late Bjorken times, $\Phi(\tau) \to \Phi_{\text{asym}}$ for $\tau \to \infty$. We concentrate on situations where this value agrees with the global minimum, $\Phi_{\text{asym}}=\Phi_\text{vac}$. [For $\theta=\pi$, there are two degenerate global minima, and $\Phi(\tau)$ dynamically selects one.]

The equation of motion for the mode functions \eqref{eq:modefunctions} simplifies in the asymptotic regime to
\begin{equation}
    \left(\partial_\tau^2+\frac{1}{\tau} \partial_\tau+\frac{k^2}{\tau^2}+{M^{\prime}}^2\right) f^\text{asym}_{k}(\tau)=0,
\label{eq:asymptoticmodeequation}
\end{equation}
where $M^\prime$ is defined in Eq.\ \eqref{eq:modM2}. By the same argument as above, the correct solution as a superposition of positive Minkowski space frequencies is given in terms of a Hankel function,
\begin{equation}
    f^\text{asym}_{k}(\tau)=\frac{\sqrt{\pi}}{2} e^{\frac{\pi k}{2}} H_{i k}^{(2)}\left(M^{\prime} \tau\right).
\end{equation}
With this, the fluctuation field becomes in the asymptotic region
\begin{equation}
    \begin{split}
    \varphi^\text{asym}(\tau, \eta) =\int &\frac{\mathrm{d} k}{2 \pi}\Big\{ a^\text{asym}_{k} f^\text{asym}_{k}(\tau) e^{i k \eta} \\ & + a^{\text{asym}\dagger}_{k} f^{\text{asym}*}_{k}(\tau) e^{-i k \eta}\Big\}.
    \end{split}
    \label{eq:phiasympmodeexpasion}
\end{equation}
To this operator corresponds a state $\left|0_{\text {asym}}\right\rangle$ that represents the Minkowski space vacuum at very late times.

Accordingly, the asymptotic particle number in Minkowski space can be determined using the asymptotic particle number operator $\hat N_k = a^{\text {asym}\dagger} a^{\text {asym}}$. We also define the occupation number as its expectation value, $n_k = \langle  \Omega |\hat N_k | \Omega \rangle$. (Recall that the state $|\Omega \rangle$ has the property $a_k^\text{init}|\Omega \rangle = 0$ but this does not imply $a_k^\text{asym}|\Omega \rangle = 0$.)

Using the usual Bogoliubov theory arguments (see, for example, Ref.~\cite{Davies, Aarts:2007qu, Traschen:1990sw, diSessa:1974ve}), one can obtain the following expression for the occupation number of the asymptotic state 
\begin{equation}
n_k = |\beta_k|^2=\tau^2 \left|f^\text{asym}_k(\tau) \partial_\tau f_k(\tau) - f_k(\tau) \partial_\tau f^\text{asym}_k(\tau)\right|^2.
\label{eq:occnumbfinal}
\end{equation}
Because $f_k^\text{asym}(\tau)$ is only a proper solution to the mode equation \eqref{eq:modefunctions} at asymptotically late time, the expression in Eq.\ \eqref{eq:occnumbfinal} only has strict physical meaning in the limit $M \tau \gg 1$.
The left panel of Fig.~\ref{fig:modefunctions2} shows the particle number $n_k$ against the wave vector $k$ at different times $\tau$ for the choice of parameters $\theta = 0$ and $\kappa=4.0$ (left panel). One observes that around $M \tau = 510$, the convergence with $M \tau$ is approaching, at least for some choices of $\theta$, $\kappa$, and $k$. The spectrum displays a peak at $k\approx0$. This is the expected behavior: excitations with small energy can be produced, while for large $k$, the kinetic term $\sim k^2$ in Eq.\ \eqref{eq:modefunctions} dominates over the time-dependent term in the second line and particle production is suppressed.

\subsection{Total particle number and temperature} \label{subsec:Temp}

\begin{figure*}
    \includegraphics[width=.49\textwidth]{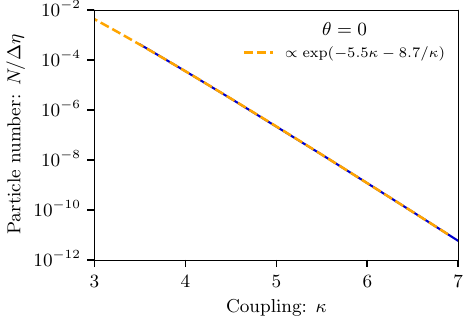}
  \includegraphics[width=.49\textwidth]{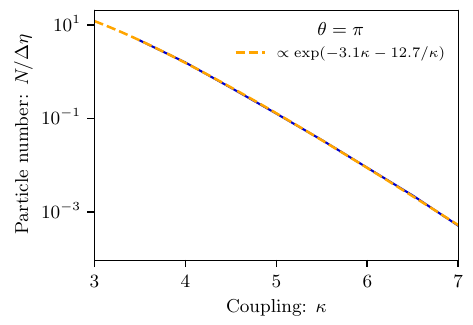}
    \caption{Total number of particles per rapidity interval in dependence on the coupling constant $\kappa$ for $\theta = 0$ (left panel) and $\theta = \pi$ (right panel). The fit has been made in the range $3.5 \leq \kappa \leq 7.0 $ for $\theta = 0, \pi$.}
    \label{fig:Temperature}
\end{figure*}

The total particle number per unit rapidity is given by the integral over occupation numbers,
\begin{equation}
\frac{N}{\Delta \eta} = \int \frac{\mathrm{d} k}{2 \pi} n_k.
\end{equation}
In Fig.~\ref{fig:Temperature}, the numerical results for the total particle number per rapidity interval $N/\Delta \eta$ are plotted as a function of the coupling $\kappa$ defined in Eq.\ \eqref{eq:dimensionlesscoupling}. We are especially interested in large values of $\kappa$ corresponding to a large fermion mass $m$.  
In this regime, for $\theta = 0$, or $\theta=\pi$, the total number of particles per rapidity interval decreases as the coupling increases. 
This behavior is expected since the oscillations of the background field around $\Phi_{\mathrm{asym}}$ are very small, and the mode functions $f$ and $f^{\mathrm{asym}}$ look very similar, which causes the difference in Eq.~\eqref{eq:occnumbfinal} to vanish.

More specifically, $N / \Delta \eta$ decays exponentially. We can make a fit to the numerical data to obtain
\begin{equation}
    \begin{aligned}
    \frac{N}{\Delta \eta} & \propto \exp \left(-5.5 \kappa- \frac{8.7}{ \kappa} \right)  \hspace{0.5cm} \text {  for  }  \theta=0,\\
    \frac{N} {\Delta \eta} & \propto \exp \left(-3.1\kappa- \frac{12.7}{ \kappa}\right) \hspace{0.5cm}  \text { for }  \theta=\pi.
    \end{aligned}
\label{eq:fittedfunction}
\end{equation}
The $1 / \kappa$-term in the exponent takes into account deviations of $N / \Delta \eta$ for smaller values of the coupling, where the field initial value deviates from the ground state more, which enhances the resonance phenomena due to its oscillations. However, for large values of $\kappa$, i.e., for highly massive or weakly coupled fermions, this term can be neglected and Eq.~\eqref{eq:fittedfunction} becomes a Boltzmann-type factor, 
\begin{equation}
   \frac{N}{\Delta \eta} \propto \exp(-\alpha_{\mathrm{fit}}\kappa) = \exp\left(-\frac{m}{T}\right),
\end{equation} 
with temperature
\begin{equation}
    T = \frac{\sqrt{2 \sigma}}{\alpha_{\mathrm{fit}}\exp(\gamma)},
\end{equation} 
proportional to the square root of the string tension $\sigma= q^2 / 2$ (cf.\ Eq.~\eqref{eq:fieldenergy_QED1+1}). 
The $\sqrt{\sigma}$-temperature dependence can also be observed phenomenologically in particle production within QCD~\cite{Becattini:2008tx}. Assuming a string tension of  $\sigma=0.19$ GeV$^2$~\cite{Becattini:2008tx}, our results for $\alpha_\text{fit}$ translate to the temperatures  
\begin{equation}
    \begin{aligned}
    T_{\theta=0} &= 63 \text{ MeV}, \\ T_{\theta=\pi}& = 112\text{ MeV},
    \end{aligned}
\end{equation}
which are of the same order of magnitude as the hadronization temperature $T_H \approx160$$-$$170$ MeV~\cite{Becattini:2008tx}.
The parameter $\theta=\pi$ seems to be closer to the actual result. However, $\theta=0$ is of the same order of magnitude, indicating that other angles may yield comparable results. Overall, it can be assumed that the numerical outcome is relatively consistent across various angles and different choices of the initialization of the fluctuations and that this result remains robust.
Some discrepancies in the results are expected since QED$_{1+1}$ is only a toy model for QCD. 
The Schwinger model offers valuable insights into confinement and thermalization mechanisms, suggesting that dimensional reduction and breaking of the confining string through tunneling present in the dynamics of the Schwinger model are key effects for the observation of the apparent thermal features. 
However, it cannot fully replicate QCD in higher dimensions. It also lacks some important features of QCD, like the gauge dynamics, multiple fermion flavors, and the non-Abelian symmetry group $SU(3)$.  

The above results have been obtained for fluctuations around a decaying false vacuum as described in Sec.~\ref{sec:Falsevacuum}. The calculation can also be done for fluctuations around the background solution described in Sec.~\ref{sec:ExpandingStringsDynamics} with a similar result, but temperatures given by $T_{\theta=0} \approx 1 \text{ GeV}, T_{\theta=\pi} \approx 2 \text{ GeV}.$
\section{Analog quantum simulator of the massive Schwinger model} \label{sec:AnalogQuantumSimulation}

This section presents an experimental proposal for an analog quantum simulation of the massive Schwinger model. An experimental realization would be compelling for checking the theoretical assumptions that the numerical predictions rely upon. In particular, it would be very insightful to observe the phenomenon of false vacuum decay and the production of particles in a laboratory through a tabletop experiment. The underlying idea is to prepare a system of ultracold atoms and adjust their interactions such that the effective potential becomes as close as possible to that of bosonized QED$_{1+1}$.    
By measuring the total particle number $N/\Delta \eta$ in the cold atom experiment, we can draw conclusions about the temperature of particles produced by vacuum decay in the Schwinger model. In the limit of large fermion mass $m$, this could describe particle production after electron-positron collisions in QED$_{1+1}$.
Here, we will briefly discuss the experimental setup and the derivation of the corresponding effective potential.

\subsection{Experimental setup}

\begin{figure*}
    \includegraphics[width=.49\textwidth]{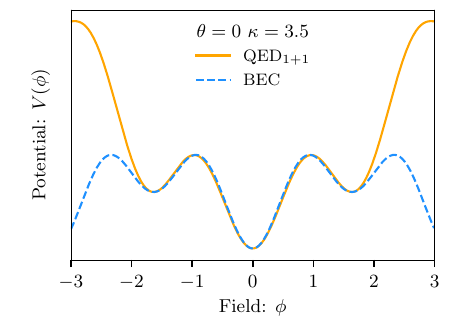}
    \includegraphics[width=.49\textwidth]{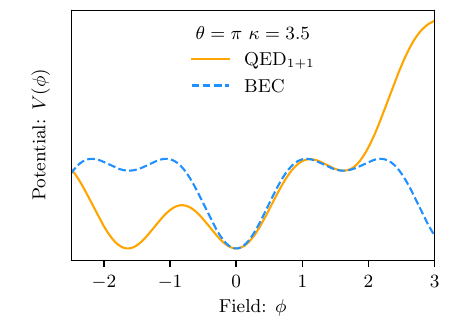}
    \caption{Comparison of the potential of the cold atom system (solid) and the potential of the massive Schwinger model (dashed) for two vacuum angles: $\theta = 0$ (left panel) and $\theta = \pi$ (right panel) and coupling ratio $\kappa=3.5$. The parameters $V_0$ and $\lambda$ of the cold atom system defined in Eq.~\eqref{eq:coldatomsystem parameters} have been chosen such that the two potentials match in the best possible way in the region between the true and false vacuum.}
    \label{fig:matching}
\end{figure*}

In Refs.~\cite{Fialko_2015, Fialko_2017, Braden:2019vsw, Braden:2017add} an experiment is proposed to test false vacuum decay using a two-component Bose-Einstein condensate made of ultracold and highly diluted atoms confined to one spatial dimension by a trap. 
One way to create a system that can simulate vacuum decay is by using atoms such as ${}^{87} \mathrm{Rb}$ or ${}^{41}\mathrm{K}$, where the condensate is made up of two different hyperfine states. Recently, a new proposal has expanded the possible experimental setups by showing that any stable mixture between two states of a bosonic isotope can be used as a relativistic analogue~\cite{Jenkins:2023npg}.
The Hamiltonian density reads
\begin{equation}
\begin{aligned}
\mathscr{H}= & -\sum_{j=1}^2\left\{\frac{1}{2 m} \partial_x \psi_j^\dagger \partial_x \psi_j+\frac{h}{2}(\psi_j^\dagger \psi_j)^2\right\} \\ & +h_c\psi_1^\dagger \psi_1\psi_2^\dagger\psi_2 -\frac{\nu}{2}\left(\psi_1^{\dagger} \psi_2+\psi_2^{\dagger} \psi_1\right).
\label{eq:Hamiltoniancoldatoms}
    \end{aligned}
\end{equation}
The two components in the system are represented by the field operators $\psi_1$ and $\psi_2$. Both have the same mass $m$, and their interactions are determined by two coupling constants, namely $h$, the coupling of each component with itself, and $h_c$, which is related to an intercomponent interaction. 
The external potential of the trap in which the atoms are confined has been omitted here. 
Additionally, an external field, e.g., a radio frequency field, induces the transition between the two states with a transition rate $\nu$.  
Furthermore, the transition rate $\nu_0$ is modulated with radio frequency $\omega$, the amplitude of the modulation being
$\nu(t)=\nu_0+ \delta \omega \cos (\omega t)$.
To bring the Hamiltonian~\eqref{eq:Hamiltoniancoldatoms} into a more convenient form, the fields are expressed in terms of their densities $\rho_j=\psi_j^\dagger \psi_j$ and phases $\vartheta_j$,
\begin{equation}
    \psi_j=\sqrt{\rho_j} e^{i \vartheta_j}, \quad j \in\{1,2\}.
\end{equation}
Additionally, we can switch to the set of variables that consists of the mean $\rho$ and the relative density $\epsilon$ defined as
    $\rho=(\rho_1+\rho_2)/2, \, \epsilon=(\rho_2-\rho_1)/2,$
and the sum $\zeta$ and the difference $\phi$ of the phases
    $\zeta=\vartheta_1+\vartheta_2, \, \phi=\vartheta_2-\vartheta_1.$
The relative phase can be measured with an interferometer~\cite{egorov2011long}.
We will focus on the relative phase $\phi$ in the following. This should not be confused with the above-mentioned scalar field $\phi$. Moreover, the frequency variation $\omega$ is assumed to be large in comparison to the timescales of the system, specifically $\omega_0=2 \sqrt{\nu_0 h \bar{n}}$ where $\bar{n}$ is the expectation value of the mean particle density. This allows time averaging and gives the final effective Lagrangian after integrating out the densities $\rho$ and $\epsilon$
\begin{equation}
    \mathcal{L}=\frac{1}{2}\left(\partial_t \phi\right)^2-\frac{c_s^2}{2} \left(\partial_x \phi\right)^2-V_0\left(-\cos\phi+\frac{\lambda^2}{2}  \sin ^2\phi\right),
    \label{eq:Lagrangiancoldatom}
\end{equation}
with parameters
    \begin{equation}
\begin{aligned}
        c_s^2 & =\frac{\left(h-h_c\right) \bar{n}\left(1-2 \delta^2\right)}{m}, \\ V_0 & =4 \nu_0 \bar{n}\left(h-h_c\right)\left(1-2 \delta^2\right), \\
        \lambda^2 & =\frac{2 \delta^2\left(h-h_c\right) \bar{n}}{\nu_0}.
\end{aligned}
\label{eq:coldatomsystem parameters}
\end{equation}
The parameter $V_0$ scales the potential in the vertical direction, whereas $\lambda$ regulates the barrier height between true and false vacuum.
The parameter $c_s$ in the Lagrangian can be interpreted as the sound speed in the given system. The shape of the potential, displayed in Fig.~\ref{fig:matching}, is the dashed curve, which is a modified version of the sine-Gordon model with an additional squared sine modulation.
\subsection{Comparison of the Potentials}
The couplings of the potential term in Eq.~\eqref{eq:Lagrangiancoldatom} can be adjusted such that it is very similar to that of the Schwinger model. Indeed, this similarity is limited to the region between the true vacuum $\phi_{TV}$ and the false vacuum $\phi_{FV}$, which is the most relevant part of the potential for vacuum decay. The experimental parameters $V_0$ and $\lambda$ of the cold atom system have been adjusted such that the shape of its potential is as close to that of the Schwinger model as possible. Moreover, the field $\phi$ needs to be scaled to match the extrema positions of the two potentials.

From the adjustment of the BEC potential for various values of the Schwinger model parameter $\kappa$, one can read off a linear mapping between $\lambda^2$ and $\kappa$ that depends on the choice of vacuum angle
\begin{equation}
\begin{array}{lll}
    \kappa=0.69 \times \lambda^2+0.48 & \text { for } & \theta=0, \\
    \kappa=1.35 \times \lambda^2+0.63 & \text { for } & \theta=\pi. \\
\end{array}
\label{eq:matching_g_and_lambda}
\end{equation}
A similar procedure as described in Sec.~\ref{sec:ParticleProduction} can be repeated to quantify the total number of particles per rapidity interval. 
By integrating each spectrum, the total particle number per rapidity interval $N/\Delta \eta$ is obtained by fitting as
\begin{equation}
    \frac{N}{\Delta \eta} \propto \exp\Big(-3.5\times \lambda^2 +1.9/\lambda^2\Big).
\end{equation}
For large $\lambda$, the first term in the exponent is dominant, and the decrease of the total particle number per rapidity interval $N/\Delta \eta$  is characterized by an exponential function with an argument proportional to $\lambda^2$. The factor in front of $\lambda^2$ agrees, up to a few percent, with that in front of $\kappa$ in equations~\eqref{eq:fittedfunction} if we use the transformation~\eqref{eq:matching_g_and_lambda}. 

In conclusion, such an experimental setup could be used to quantum simulate important aspects of the massive Schwinger model. 

\vspace{-.9em}
\section{Conclusions} \label{sec:Conclusions}
In this work we studied the dynamics and particle production from a string between a quark-antiquark pair produced, for example, in an electron-positron collision. As a model, we used the bosonized version of the massive Schwinger model, which makes dynamics particularly easy to simulate numerically. We examine the dynamics of the coherent field and the 
particle production. 
Remarkably, the total particle number per rapidity interval to the fermion mass for highly massive or weakly coupled fermions turns out to be well represented by a Boltzmann factor. 

This finding confirms previous investigations where an expanding string was found at early times to be governed locally by a time-dependent temperature as a result of longitudinal entanglement \cite{Berges:2017zws, Berges:2017hne, Berges:2018cny}. The setup of the present work is more complete and the time-dependent temperature $T=\hbar/(2\pi \tau)$ of Ref.\ \cite{Berges:2017zws} gets replaced dynamically by a constant value proportional to the string tension scale $\sqrt{\sigma}$ for the dependence of asymptotic particle number per unit rapidity as a function of the fermion mass $m$. 
The precise value of this hadronization temperature can be determined through a fitting procedure. Quantitatively, it is reasonably close to the phenomenologically determined value. 

We also outlined a possible analog quantum simulator of the massive Schwinger model. 
It would be very insightful to observe the phenomenon of false vacuum decay and the production of particles in a laboratory through a tabletop experiment, such that the effective experimental potential becomes as close as possible to that of bosonized QED$_{1+1}$.    
By measuring the total particle number $N/\Delta \eta$ in the cold atom experiment, one can draw conclusions about the temperature of particles produced by vacuum decay in the Schwinger model. In the limit of large couplings $\lambda$ and $\kappa$, i.e., large fermion mass $m$, this could describe particle production after electron-positron collisions in QED$_{1+1}$.

Very interesting extensions of this work include the consideration of flavor and color numbers, the incorporation of the production description beyond mesons, and taking into account baryons. Moreover, for the bounce scenario discussed in Sec.~\ref{sec:Falsevacuum}, the precise determination of the quantum state of fluctuations after the bounce would be highly valuable. The question of whether this can be achieved through analytic continuation from a calculation in Euclidean space is left for future work.
\subsection*{ACKNOWLEDGEMENTS}
 For discussions on related topics, the authors would like to thank Sebastian Erne, Michael Heinrich, Jörg Schmiedmayer, and Raju Venugopalan.
L.B., J.B., and S.F. acknowledge support by the DFG under the Collaborative Research Center SFB
1225 ISOQUANT (Project-ID 27381115) and the Heidelberg STRUCTURES Excellence Cluster under Germany’s Excellence Strategy EXC2181/1-390900948.

\appendix \label{Appendix}
\section{False vacuum decay} \label{App:FalseVacuumDecay}
This appendix completes Sec.~\ref{sec:Falsevacuum}, particularly explaining how to calculate the bounce solution $\phi_B$. 
The Euclidean action corresponding to~\eqref{eq:MassiveSchwingeraction} is obtained by integrating along the imaginary Minkowski time, $t=-iy$, with the Euclidean time $y$, and by changing the overall sign of the action,
\begin{equation}
    S_E=\int \mathrm{d} y \mathrm{d} x\left(\frac{1}{2}\left(\partial_{y} \phi\right)^2 + \frac{1}{2}\left(\partial_x \phi\right)^2+V(\phi)\right), 
\end{equation}
We define the so-called bounce solution $\phi_B$, which solves the classical Euclidean equation of motion 
\begin{equation}
    \partial_{y}^2 \phi+\partial_x^2 \phi=M^2 \phi + M^2 \kappa \sin \left(2 \sqrt{\pi} \phi+\theta\right).\label{eq:bounceequation}
\end{equation}
The boundary conditions read $\phi_B\left(y, x\right) \rightarrow \phi_{FV}$ for $y \rightarrow \pm \infty ,$
and 
$\phi_B\left(y, x\right) \rightarrow \phi_{FV}$ for $|x| \rightarrow \infty$. 
It can be shown that the action $S_E(\phi_B)$ is minimized by a spherically symmetric bounce solution~\cite{Coleman:1977py}. Therefore, we only consider spherical symmetric solutions of~\eqref{eq:bounceequation} and introduce the radial coordinate $r^2 = y^2 + x^2$, then
\begin{equation}
    \partial_r^2 \phi_B+\frac{1}{r} \partial_r \phi_B= M^2 \Big( \phi + \kappa \sin(2 \sqrt{\pi} \phi + \theta) \Big).
\end{equation}
Boundary conditions $\phi_B \rightarrow \phi_{FV}$ for $r \rightarrow \infty$
together $\partial_r \phi_B(r=0)=0$ have to be fulfilled to avoid singularities. The shooting method is typically applied to solve this equation numerically. In this method, one specifies the initial value of $\phi_B$ at $r=0$ to ensure the bounce approaches the false vacuum at the spatial boundaries.

\section{Renormalization} \label{app:renormalization_Schwinger_model}
This appendix completes Sec.~\ref{sec:QED1+1}, specifically providing explanation based on the functional renormalization group method to justify the renormalized potential (Eq.~\eqref{eq:renormalizedPotential})
The theory in Eq.~\eqref{eq:QED1+1Lagrangian} is alternatively described by the microscopic action
\begin{equation}
    S_\Lambda=\int \mathrm{d}t\mathrm{d}x \left\{-\frac{1}{2} \eta^{\mu \nu} \partial_\mu \phi \partial_\nu \phi-V_\Lambda(\phi)\right\},
\label{eq:MassiveSchwingeraction}
\end{equation}
with the scalar potential
\begin{equation}
    V_\Lambda(\phi)=\frac{1}{2}M^2 \phi^2 - u_\Lambda \cos (2 \sqrt{\pi} \phi+\theta),
\label{eq:bosonicpotential}
\end{equation}
where $M=q/\sqrt{\pi}$ denotes the Schwinger mass. This classical action is associated with a corresponding ultraviolet (UV) regulator scale that we take as a sharp momentum cutoff $\Lambda$.
The parameter $u_\Lambda$ in a functional integral prescription is also proportional to $\Lambda$ and the fermion mass $m$ as \cite{Jentsch:2021trr} 
\begin{eqnarray}
    u_\Lambda = \frac{\exp(\gamma)}{2\pi} \Lambda m,
\end{eqnarray}
with $\gamma\approx 0.5772$ the Euler-Mascheroni constant.
This cutoff dependence of the coupling is, at first sight, surprising, given that no such regulator dependence is shown by the fermionic formulation in Eq.~\eqref{eq:QED1+1Lagrangian}. Due to the equivalence with the fermionic formulation, one expects that all final physical observables depend only on $m$ and $q$ and become independent of $\Lambda$. In practice, we are interested in the properties of the theory at macroscopic or infrared (IR) scales $k$. Indeed, in the next subsection, we show that a renormalization procedure can eliminate the dependence on the UV regularization.

The bosonized theory in Eq.\ \eqref{eq:MassiveSchwingeraction} is an interacting quantum field theory subject to renormalization. 
A detailed discussion of renormalization in the functional formulation, addressing states with zero and nonzero temperature, can be found in Ref.\ \cite{Jentsch:2021trr}. The path integral systematically includes quantum fluctuations into correlation functions when moving from UV to IR, where the scale-dependent fluctuations enter corresponding coupling constants of the scale-dependent effective action $\Gamma_k$.  In the limit $k\gg \Lambda$ all quantum fluctuations are suppressed, such that $\Gamma_k[\phi] = S_\Lambda[\phi]$. On the other side, for $k\to 0$, the flowing action would approach the full quantum effective action $\Gamma_{k=0} [\phi] =\Gamma[\phi]$, see Ref.~\cite{Jentsch:2021trr} for a more detailed discussion.
The regulator function is chosen such that $\Gamma_k$ smoothly interpolates between the microscopic action $S_\Lambda$ for $k=\Lambda$ and the macroscopic action $\Gamma$ for $k=0$.
We assume that $\Gamma_k[\phi]$ is of the same form as Eq.~\eqref{eq:MassiveSchwingeraction} with potential term as in Eq.~\eqref{eq:bosonicpotential}, except for the replacement $u_\Lambda \to u_k$,
whereas the derivative term keeps its bare form. 
For the purpose of the present work, we shall be satisfied with this simplified approach where only a single parameter, the coupling $u_{\Lambda}$, gets replaced by the renormalized coupling $u$. There is, in fact, a good reason why $M$ does not need to be renormalized, which is explained below.
The $k$-dependent effective potential $V_{k}(\phi)$ satisfies the exact equation~\cite{Wetterich:1992yh},

\begin{equation}
\begin{aligned}
    \partial_k V_k(\phi) = &   \frac{1}{2} \int\limits_{p^2<\Lambda^2} \frac{\mathrm{d}^2p}{(2\pi)^2} \frac{2k}{p^2+k^2+V_k^{(2)}(\phi)} \\
    = & \frac{k}{4\pi} \ln\left( \frac{\Lambda^2+k^2+V_k^{(2)}(\phi)}{k^2 + V_k^{(2)}(\phi)} \right),
\end{aligned}
\end{equation}
with
\begin{equation}
    V_k^{(2)}(\phi) \equiv \frac{\partial^2 V_k(\phi)}{\partial\phi^2} = M^2 + 4\pi  u_k \cos(2\sqrt{\pi}\phi + \theta).
\end{equation}
We have chosen a simple mass-like infrared regulator $R_k(p)=k^2$ and applied a sharp ultraviolet momentum cutoff $\Lambda$ for consistency.

Note that the flow $\partial_k V_k(\phi)$ is invariant under field transformations $\phi \to \phi + \sqrt{\pi}$ because this is the case for $V_k^{(2)}(\phi)$. Accordingly, a periodicity-breaking term as $M^2\phi^2/2$ in Eq.\ \eqref{eq:bosonicpotential}, does not get renormalized. 

To obtain a flow equation for $u_k$, we use the inverse Fourier expansion scheme,
\begin{equation}
    \partial_k u_k = -\frac{2}{\sqrt{\pi}} \int_0^{\sqrt{\pi}} \mathrm{d}\phi \cos(2\sqrt{\pi}\phi+\theta) \partial_k V_k(\phi).
\end{equation}
We concentrate furthermore on the leading term in an expansion in $u_k$ and find
\begin{equation}
    \partial_k u_k = - u_k k \left[ \frac{1}{\Lambda^2 + k^2 + M^2} - \frac{1}{k^2 + m^2} \right].
\end{equation}
This can be integrated with respect to $k$ to yield
\begin{equation}
    \ln \left(\frac{u_k}{u_\Lambda}\right) = - \frac{1}{2} \ln\left( \frac{\Lambda^2 + k^2 + M^2}{k^2 + M^2} \right).
\end{equation}
We have chosen an integration constant such that $u_k = u_\Lambda$ for $k \gg \Lambda$. From this we find at $k=0$ and for $\Lambda\gg M$
\begin{equation}
    u = u_0 = u_\Lambda \frac{M}{\Lambda}.
\end{equation}
The final form of the renormalized potential $V(\phi)$ is given by Eq.~\eqref{eq:renormalizedPotential}.

    \nocite{*}
\bibliography{bibliographyhadro}
\end{document}